\documentclass[twocolumn,showpacs,preprintnumbers,superscriptaddress,amsmath,amssymb,PRL]{revtex4}
\usepackage{graphicx}
\usepackage{dcolumn}
\usepackage{float}
\usepackage{mathptmx, courier, pifont}
\usepackage[scaled=0.92]{helvet}
\usepackage[T1]{fontenc}
\usepackage{textcomp}
\usepackage{color}

\usepackage[dvipsnames]{xcolor}
\usepackage[colorlinks=true,urlcolor=Blue,linkcolor=Blue]{hyperref}
\usepackage[all]{hypcap}

\begin{document}


\title{The Masses of the $T_z=-3/2$ Nuclei $^{27}$P and $^{29}$S}

\author{C. Y. Fu}
\affiliation{Key Laboratory of High Precision Nuclear Spectroscopy and Center for Nuclear Matter Science, Institute of Modern Physics, Chinese Academy of Sciences, Lanzhou 730000, People's Republic of China}
\affiliation{University of Chinese Academy of Sciences, Beijing 100049, People's Republic of China}
\affiliation{Lanzhou University, Lanzhou 730000, People's Republic of China}
\author{Y.~H.~Zhang}
\affiliation{Key Laboratory of High Precision Nuclear Spectroscopy and Center for Nuclear Matter Science, Institute of Modern Physics, Chinese Academy of Sciences, Lanzhou 730000, People's Republic of China}
\author{X.~H.~Zhou}\thanks{Corresponding author. Email address: zxh@impcas.ac.cn}
\affiliation{Key Laboratory of High Precision Nuclear Spectroscopy and Center for Nuclear Matter Science, Institute of Modern Physics, Chinese Academy of Sciences, Lanzhou 730000, People's Republic of China}
\author{M.~Wang}\thanks{Corresponding author. Email address: wangm@impcas.ac.cn}
\affiliation{Key Laboratory of High Precision Nuclear Spectroscopy and Center for Nuclear Matter Science, Institute of Modern Physics, Chinese Academy of Sciences, Lanzhou 730000, People's Republic of China}
\author{Yu.~A. Litvinov}\thanks{Corresponding author. Email address: y.litvinov@gsi.de}
\affiliation{Key Laboratory of High Precision Nuclear Spectroscopy and Center for Nuclear Matter Science, Institute of Modern Physics, Chinese Academy of Sciences, Lanzhou 730000, People's Republic of China}
\affiliation{GSI Helmholtzzentrum f\"{u}r Schwerionenforschung,
	Planckstra{\ss}e 1, 64291 Darmstadt, Germany}
\author{K.~Blaum}
\affiliation{Max-Planck-Institut f\"{u}r Kernphysik, Saupfercheckweg 1, 69117 Heidelberg, Germany}
\author{H.~S.~Xu}
\affiliation{Key Laboratory of High Precision Nuclear Spectroscopy and Center for Nuclear Matter Science, Institute of Modern Physics, Chinese Academy of Sciences, Lanzhou 730000, People's Republic of China}

\author{X.~Xu}
\affiliation{Key Laboratory of High Precision Nuclear Spectroscopy and Center for Nuclear Matter Science, Institute of Modern Physics, Chinese Academy of Sciences, Lanzhou 730000, People's Republic of China}
\author{P.~Shuai}
\affiliation{Key Laboratory of High Precision Nuclear Spectroscopy and Center for Nuclear Matter Science, Institute of Modern Physics, Chinese Academy of Sciences, Lanzhou 730000, People's Republic of China}
\author{Y.~H.~Lam}
\affiliation{Key Laboratory of High Precision Nuclear Spectroscopy and Center for Nuclear Matter Science, Institute of Modern Physics, Chinese Academy of Sciences, Lanzhou 730000, People's Republic of China}
\author{R.~J.~Chen}
\affiliation{Key Laboratory of High Precision Nuclear Spectroscopy and Center for Nuclear Matter Science, Institute of Modern Physics, Chinese Academy of Sciences, Lanzhou 730000, People's Republic of China}
\author{X.~L.~Yan}
\affiliation{Key Laboratory of High Precision Nuclear Spectroscopy and Center for Nuclear Matter Science, Institute of Modern Physics, Chinese Academy of Sciences, Lanzhou 730000, People's Republic of China}
\author{T.~Bao}
\affiliation{Key Laboratory of High Precision Nuclear Spectroscopy and Center for Nuclear Matter Science, Institute of Modern Physics, Chinese Academy of Sciences, Lanzhou 730000, People's Republic of China}
\author{X.~C.~Chen}
\affiliation{Key Laboratory of High Precision Nuclear Spectroscopy and Center for Nuclear Matter Science, Institute of Modern Physics, Chinese Academy of Sciences, Lanzhou 730000, People's Republic of China}
\affiliation{Max-Planck-Institut f\"{u}r Kernphysik, Saupfercheckweg 1, 69117 Heidelberg, Germany}
\author{H.~Chen}
\affiliation{Key Laboratory of High Precision Nuclear Spectroscopy and Center for Nuclear Matter Science, Institute of Modern Physics, Chinese Academy of Sciences, Lanzhou 730000, People's Republic of China}
\affiliation{Graduate University of Chinese Academy of Sciences,
	Beijing, 100049, People's Republic of China}
\author{J.~J.~He}
\affiliation{Key Laboratory of Optical Astronomy, National Astronomical Observatories, Chinese Academy of Sciences, Beijing, 100012, People's Republic of China}
\affiliation{Key Laboratory of High Precision Nuclear Spectroscopy and Center for Nuclear Matter Science, Institute of Modern Physics, Chinese Academy of Sciences, Lanzhou 730000, People's Republic of China}
\author{S.~Kubono}
\affiliation{Key Laboratory of High Precision Nuclear Spectroscopy and Center for Nuclear Matter Science, Institute of Modern Physics, Chinese Academy of Sciences, Lanzhou 730000, People's Republic of China}
\author{D.~W.~Liu}
\affiliation{Key Laboratory of High Precision Nuclear Spectroscopy and Center for Nuclear Matter Science, Institute of Modern Physics, Chinese Academy of Sciences, Lanzhou 730000, People's Republic of China}
\affiliation{Graduate University of Chinese Academy of Sciences,
	Beijing, 100049, People's Republic of China}
\author{R.~S.~Mao}
\affiliation{Key Laboratory of High Precision Nuclear Spectroscopy and Center for Nuclear Matter Science, Institute of Modern Physics, Chinese Academy of Sciences, Lanzhou 730000, People's Republic of China}
\author{X.~W.~Ma}
\affiliation{Key Laboratory of High Precision Nuclear Spectroscopy and Center for Nuclear Matter Science, Institute of Modern Physics, Chinese Academy of Sciences, Lanzhou 730000, People's Republic of China}
\author{M.~Z.~Sun}
\affiliation{Key Laboratory of High Precision Nuclear Spectroscopy and Center for Nuclear Matter Science, Institute of Modern Physics, Chinese Academy of Sciences, Lanzhou 730000, People's Republic of China}
\affiliation{Graduate University of Chinese Academy of Sciences,
	Beijing, 100049, People's Republic of China}
\author{X.~L.~Tu}
\affiliation{Key Laboratory of High Precision Nuclear Spectroscopy and Center for Nuclear Matter Science, Institute of Modern Physics, Chinese Academy of Sciences, Lanzhou 730000, People's Republic of China}
\affiliation{Max-Planck-Institut f\"{u}r Kernphysik, Saupfercheckweg 1, 69117 Heidelberg, Germany}
\author{Y.~M.~Xing}
\affiliation{Key Laboratory of High Precision Nuclear Spectroscopy and Center for Nuclear Matter Science, Institute of Modern Physics, Chinese Academy of Sciences, Lanzhou 730000, People's Republic of China}
\affiliation{Graduate University of Chinese Academy of Sciences,
	Beijing, 100049, People's Republic of China}
\author{P.~Zhang}
\affiliation{Key Laboratory of High Precision Nuclear Spectroscopy and Center for Nuclear Matter Science, Institute of Modern Physics, Chinese Academy of Sciences, Lanzhou 730000, People's Republic of China}
\affiliation{Graduate University of Chinese Academy of Sciences,
	Beijing, 100049, People's Republic of China}
\author{Q.~Zeng}
\affiliation{Research Center for Hadron Physics, National Laboratory of Heavy Ion Accelerator Facility in Lanzhou and University of Science and Technology of China, Hefei 230026, People's Republic of China}
\affiliation{Key Laboratory of High Precision Nuclear Spectroscopy and Center for Nuclear Matter Science, Institute of Modern Physics, Chinese Academy of Sciences, Lanzhou 730000, People's Republic of China}
\author{X.~Zhou}
\affiliation{Key Laboratory of High Precision Nuclear Spectroscopy and Center for Nuclear Matter Science, Institute of Modern Physics, Chinese Academy of Sciences, Lanzhou 730000, People's Republic of China}
\affiliation{Graduate University of Chinese Academy of Sciences,
	Beijing, 100049, People's Republic of China}

\author{W.~L.~Zhan}
\affiliation{Key Laboratory of High Precision Nuclear Spectroscopy and Center for Nuclear Matter Science, Institute of Modern Physics, Chinese Academy of Sciences, Lanzhou 730000, People's Republic of China}
\author{S.~Litvinov}
\affiliation{GSI Helmholtzzentrum f\"{u}r Schwerionenforschung,
	Planckstra{\ss}e 1, 64291 Darmstadt, Germany}
\author{G.~Audi}
\affiliation{CSNSM-IN2P3-CNRS, Universit\'{e} de Paris Sud, F-91405
	Orsay, France}
\author{T.~Uesaka}
\affiliation{RIKEN Nishina Center, RIKEN, Saitama 351-0198, Japan}
\author{Y. Yamaguchi}
\affiliation{RIKEN Nishina Center, RIKEN, Saitama 351-0198, Japan}
\author{T. Yamaguchi}
\affiliation{Department of Physics, Saitama University, Saitama 338-8570, Japan}
\author{A.~Ozawa}
\affiliation{Insititute of Physics, University of Tsukuba, Ibaraki 305-8571, Japan}
\author{B.~H.~Sun}
\affiliation{School of Physics and Nuclear Energy Engineering,
	Beihang University, Beijing 100191, People's Republic of China}
\author{Y.~Sun}
\affiliation{Department of Physics and Astronomy, Shanghai Jiao Tong University,
	Shanghai 200240, People's Republic of China}
\author{F.~R.~Xu}
\affiliation{State Key Laboratory of Nuclear Physics and Technology, School of Physics, Peking University, Beijing 100871, People's Republic of China}



\date{\today}

\begin{abstract}
Isochronous mass spectrometry has been applied in the storage ring CSRe to measure the masses of the $T_z=-3/2$ nuclei $^{27}$P and $^{29}$S.
The new mass excess value $ME$($^{29}$S) $=-3094(13)$~keV is 66(52)~keV larger than the result
of the previous $^{32}$S($^3$He,$^{6}$He)$^{29}$S reaction measurement in 1973
and a factor of 3.8 more precise.
The new result for $^{29}$S, together with those of the $T=3/2$ isobaric analog states (IAS) in $^{29}$P, $^{29}$Si, and $^{29}$Al,
fit well into the quadratic form of the Isobaric Multiplet Mass Equation IMME.
The mass excess of $^{27}$P has been remeasured to be $ME(^{27}$P$)=-685(42)$ keV.
By analyzing the linear and quadratic coefficients of the IMME in the $T_z=-3/2$ $sd$-shell nuclei,
it was found that the ratio of the Coulomb radius parameters is $R\approx0.96$ and is nearly the same for all $T=3/2$ isospin multiplets.
Such a nearly constant $R$-value, apparently valid for the entire light mass region with $A>9$, can be used to set stringent constraints on the isovector and isotensor
components of the isospin non-conserving forces in theoretical calculations.
\end{abstract}


\maketitle

The atomic mass is a fundamental property of an atomic nucleus.
Dependent on the precisions achieved, mass measurements play an important role in many areas of subatomic physics
ranging from nuclear structure and astrophysics to fundamental interactions and symmetries~\cite{Lunney03,Blaum06,100yms}.
About half a century ago, it was noted that masses of the members of an isospin multiplet should follow the quadratic equation~\cite{Wigner57,Weinberg}:
\begin{eqnarray}\label{IMME equ}
ME(\alpha,T,T_{z}) = a(\alpha,T)+b(\alpha,T)T_{z}+c(\alpha,T)T_{z}^{2},
\end{eqnarray}
where $ME$ is the mass excess of a multiplet member, \emph{T} the total isospin, $T_z=(N-Z)/2$ the isospin projection on $z$ axis,
\emph{a, b, c} parameters dependent on $\alpha(A,J^\pi,...)$.
The latter depends on the atomic mass number $A$, the spin-parity $J^\pi$, and other quantum numbers.
This local mass relation is usually called the isobaric multiplet mass equation (IMME).
Because IMME is a fundamental prediction following from the concept of isospin in nuclear physics,
any breakdown of its quadratic form would indicate that a higher-order
perturbation and/or some other mechanisms of isospin-symmetry breaking should be explored~\cite{Bentley07}.
The validity of IMME has been thoroughly checked in the $T \ge 3/2$ isospin multiplets in the $A=10\sim 60$ mass region,
see Refs.~\cite{1979IMME,2013LA29,2014MA56}.
Usually one adds an extra term, such as $dT_z^3$ and/or $eT_z^4$, to Eq.~(1) in order to measure the extent of breakdown of the quadratic IMME.
The $d$ coefficients have been systematically studied by fitting the energies of the isobaric multiplets, see Refs.~\cite{1979IMME,2013LA29,2014MA56}.
One shall also note, that the IMME is regarded as a precise tool to predict the ground-state masses
of very neutron-deficient nuclides or the excitation energies
of isobaric analog states (IAS).

For the isobaric quartets and quintets in the $sd$-shell, it has been shown~\cite{2014MA56},
that the $d$ coefficients are compatible with zero within 3$\sigma$, except for the $A=35$ quartet and the $A=32$ quintet,
for which the $d$ coefficients significantly deviate from zero by $8\sigma$ and $9\sigma$, respectively.
Furthermore, large error bars existed for the $A=29$ and $A=31$ quartets~\cite{2014MA56},
which were due to the experimental mass uncertainties of $^{29}$S~\cite{1973BE09} and $^{31}$Cl~\cite{1977BE09}.
The large uncertainty for the $A=31$ quartet has been recently removed through a
precision mass measurement of $^{31}$Cl
by the JYFLTRAP double-Penning-trap mass spectrometer~\cite{2016KA15}.
As a consequence, $^{31}$Cl was found to be 36 keV less bound with respect to the earlier measurements~\cite{1977BE09,AME2012},
and $d=-3.5\pm 1.1$~keV was obtained which deviates from zero by $3.2\sigma$.
This substantial breakdown of the quadratic IMME for $A=31$ quartet is most likely due to
the isospin mixing of the isobaric analog states in $^{31}$S as well as in $^{31}$P~\cite{2016BE19}.
As for the $A=29$ quartet, the $ME$ values of $^{29}$P, $^{29}$Si, and $^{29}$Al are known
with precisions better than 5 keV,
whereas a 50-keV mass uncertainty~\cite{1973BE09} is tabulated for $^{29}$S
in the latest Atomic-Mass Evaluation AME$^{\prime}$16~\cite{AME2016}.
The mass excess $ME(^{29}$S$)=-3160(50)$~keV
is deduced from the $^{32}$S$(^3$He,$^6$He$)^{29}$S transfer reaction performed in 1970s at Cyclotron Laboratory of Michigan State University with Spectrograph.
This large uncertainty leads to $d=9.7\pm 8.8$~keV, thus evidently calling for a
more precise mass value of $^{29}$S.

The isochronous mass spectrometry (IMS) technique established in the Cooler Storage Ring (CSR) at the Heavy Ion Research Facility in Lanzhou (HIRFL)~\cite{Xiajiawen} allows for performing precision mass measurements on short-lived nuclides~\cite{IMS,Tuxiaolin}.
Numerous results have been obtained in the past few years~\cite{Hushan13,zyh2016a}.
A relative mass precision up to $1\sim 2 \times 10^{-7}$ ($\sigma=5\sim10$ keV)
has been achieved in the latest mass measurement for the $fp$-shell nuclides~\cite{2016XU10,Zhangp16}.
In this Work, we report the direct mass measurement of the short-lived $^{29}$S
nuclide with half-life $T_{1/2}=188(4)$~ms~\cite{AME2016,1979Vi01}.


The experiment was conducted at the HIRFL-CSR acceleration complex~\cite{Xiajiawen}, which consists of a separated sector cyclotron (SSC, $K=450$),
a sector-focusing cyclotron (SFC, $K=69$), a main cooler-storage ring (CSRm) operating as a heavy-ion synchrotron, and an experimental storage ring CSRe.
The two storage rings are connected by an in-flight fragment separator RIBLL2~\cite{zwl2010}.

In the present experiment, a 468~MeV/u $^{58}$Ni$^{19+}$ primary beam of about $8\times 10^7$ particles per spill
was fast-extracted from the CSRm and focused upon a $\sim $15~mm thick $^9$Be target placed in front of the RIBLL2.
At this relativistic energy, the reaction products from the projectile fragmentation of $^{58}$Ni were emerged from the target mainly as bare nuclei.
They were then selected and analyzed~\cite{Geis92} by the RIBLL2.
Finally, a cocktail beam including the ions of interest was injected into the CSRe,
which was tuned into the isochronous ion-optical mode~\cite{Tuxiaolin,Haus00} with the transition point at $\gamma_t=1.400$.
The primary beam energy was selected according to the LISE++ simulations~\cite{Tar08} such that the $^{44}$Cr$^{24+}$ ions
had the most probable velocity with $\gamma=\gamma_t$ at the exit of the target.
Both RIBLL2 and CSRe were set to a fixed magnetic rigidity of $B\rho=5.5778$~Tm
to allow for an optimal transmission of the $T_z=-2$ nuclides centered at $^{44}$Cr.
We note, that the projectile fragments of $^{58}$Ni had a broad momentum distributions of a few percent.
All nuclides within the $B\rho$ acceptance of $\pm 0.2$\% of the RIBLL2-CSRe system have been transmitted and stored in the CSRe.
\par
The revolution times of the stored ions were measured
with a timing detector~\cite{Meibo}.
A 19 $\mu$g/cm$^2$ carbon foil is installed inside the CSRe aperture.
Each time an ion passed through the foil, secondary electrons were released from its surface and transmitted
to a micro-channel plate (MCP) counter.
The signals from the MCP were recorded without amplification by a fast digital oscilloscope.
The typical rising time of the signals was $250\sim 500$ ps~\cite{Meibo}.
\par
The revolution frequencies of the stored ions were about 1.6 MHz.
The time resolution of the ToF detector was about 50 ps.
The detection efficiency varied from $\sim 7$\% to $\sim 80$\% dependending on the charge and the overall number of stored ions
(see Refs.~\cite{Tuxiaolin,Meibo} for more details).
For each injection, a measurement time of  $300$~$\mu$s, triggered by the start pulse of the CSRe injection kicker,
was acquired by the oscilloscope, which corresponds to about $500$ revolutions of the ions in the CSRe.
In total data from 29600 injections were measured in the experiment.
The present experiment aimed at the very neutron-deficient nuclei with production yields much lower than in our previous measurements.
In order to increase the sensitivity of the measurements, only ions that circulated for more than 100 $\mu s$ and created more than 30 signals were considered.
Under these conditions, about 3$\sim$6 stored ions were identified in each injection.
This is different from previous analyses~\cite{Tuxiaolin,Ni53a,Yan13,Co51,TuPRL},
where the minimum storage time of 186 $\mu s$ within a measurement time of 200 $\mu s$ was required.
Thus, the number of ions used in the analysis could be increased.
The revolution time spectrum and the corresponding isotope identification were obtained as described in Refs.~\cite{Tuxiaolin,magnetic Cor}.

\begin{figure}[htbp]
	\includegraphics[width=1.0\columnwidth]{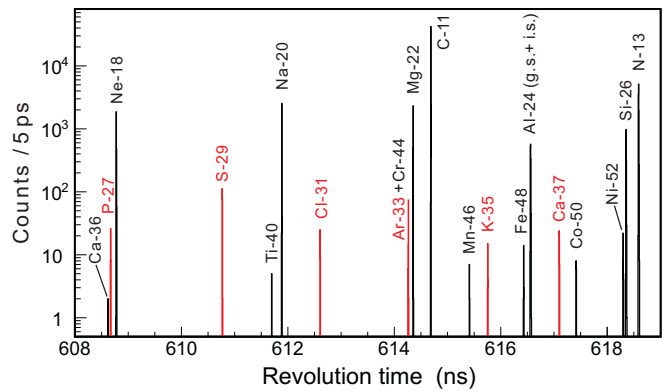}
	\caption{(Color online). Part of the measured revolution time spectrum.
	$T_{z}= -3/2$ nuclei are indicated with red colour.}
  \label{spectrum}
\end{figure}

Figure~\ref{spectrum} illustrates a part of the revolution time spectrum
zoomed in a time window of 608 ns $\leq$ \emph{t} $\leq$ 619 ns,
from which the mean revolution time $T$ and the corresponding standard deviation RMS for each peak have been determined.
The latter
exhibit a parabolic shape ranging from 1 ps to 4 ps
with a minimum at around $^{22}$Mg.
Due to a large $B\rho$ acceptance of $\pm 0.2$\% of the CSRe, the nuclides with $-1/2\le T_z\le -2$ were injected and stored in CSRe (see Fig.~\ref{spectrum}).
The counting statistics of $T_z=-2$ nuclei was relatively low and required a dedicated analysis which will be published elsewhere.

Most of the nuclides in Fig.~\ref{spectrum} have well-known masses.
Their $ME$ values from AME$^{\prime}$16~\cite{AME2016} were used
to fit their mass-to-charge ratios \emph{m/q} versus the corresponding revolution times $T$
employing a third order polynomial function.
Both, the errors of the revolution times $T$ and of the literature mass values in AME$^\prime$16 have been considered.
The mass calibration has been checked by re-determining the $ME$ values of each of the $N_c$ reference nuclides ($N_c=10$)
using the other $N_c-1$ ones as calibrants.
The normalized $\chi_{n}$ defined as:
\begin{eqnarray}\label{Chi-square equ}	\chi_{n}=\sqrt{\frac{1}{N_c}\sum\limits_{i=1}^{N_c}\frac{[(\frac{m}{q})_{i,exp}-(\frac{m}{q})_{i,AME}]^{2}}{[\sigma_{exp}(\frac{m}{q})_{i}]^{2}+[\sigma_{AME}(\frac{m}{q})_{i}]^{2}}}~~,
\end{eqnarray}
was found to be $\chi$$_{n}=1.14$.
This value is within the expected range of $\chi_{n}=1\pm0.22$ at $1\sigma$ confidence level,
indicating that no additional systematic error has to be considered.

\begin{figure}[t]
	\includegraphics[width=0.9\columnwidth]{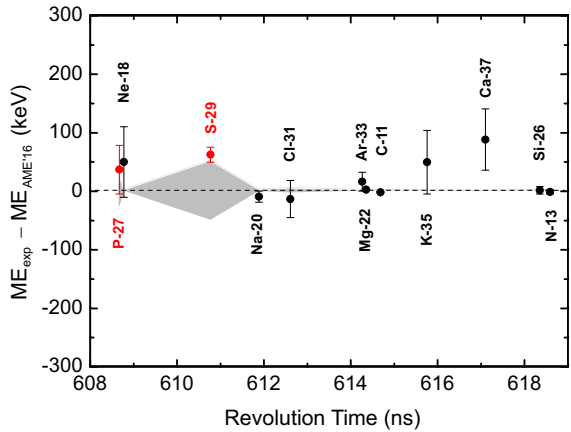}
	\caption{(Color online). Differences between $ME$ values determined in this work and from the AME$^{\prime}$16~\cite{AME2016}.
	The mass value of each of the 10 reference nuclides (black colour)
	was re-determined here by using the other 9 nuclides as references.
	The masses of $^{27}$P and $^{29}$S (red colour) were determined by using all 10 reference nuclides.
	The gray shadow indicates the 1$\sigma$ uncertainty from AME$^{\prime}$16.}
  \label{mutualcal}
\end{figure}

Figure~\ref{mutualcal} presents the differences between $ME$ values determined within this work and their literature values~\cite{AME2016}.
Our $ME$ values are in excellent agreement with AME$^{\prime}$16~\cite{AME2016}.

An interesting result from this work concerns the mass of $^{29}$S.
Our experiment yields $ME(^{29}\rm {S})=-3094(13)$~keV.
This value differs by $1.32\sigma$ from the recommended 
value in AME$^{\prime}$16 and has a precision improved by a factor of 3.8.
Our result shows that $^{29}$S is 66(52) keV less bound than the value given in Ref.~\cite{1973BE09} and in AME$^{\prime}$16~\cite{AME2016}.

Also $ME(^{27}{\rm P})=-685(42)$ keV is obtained in this work.
The literature value is $ME(^{27}{\rm P})=-722(26)$ keV and is an average of two independent experiments \cite{AME2016}.
Our value is in excellent agreement with the more recent $Q$-value
measurement of $^{28}$Si$(^7$Li,$^8$He$)^{27}$P reaction \cite{27PMSU}
from which $ME(^{27}{\rm P})=-683(41)$ keV has been deduced by
using the new value of $ME(^8$He$)= 31609.68(9)$ keV \cite{AME2016}
and disagrees by 1.24$\sigma$ to $ME(^{27}{\rm P})=-753(35)$ keV
from an older $^{32}$S$(^3$He,$^8$Li$)^{27}$P measurement \cite{1977BE09}.
The latter value was questionned in \cite{27PMSU}.
In the following we use our value and note that the conclusions do not change if an averaged value is taken.

The ground states of $^{29}$Al $(T_z=+3/2)$, $^{29}$Si $(T_z=+1/2)$, and $^{29}$P $(T_z=-1/2)$
have been previously measured with high precision~\cite{AME2016,AME2016(I),2005Si29,2015P29},
and the excitation energies of the $T=3/2$ IASs in $^{29}$Si and $^{29}$P are known~\cite{AME2016}.
The excitation energy for the $J^\pi= 5/2^+$, $T = 3/2$ IAS in $^{29}$P
was measured through proton resonance spectroscopy \cite{29P1}, $\gamma$ decay \cite{29P2}
and $\beta^+$-delayed protons from $^{29}$S \cite{29P3}.
These experimental data agree within one standard deviation.
The proton resonance energy and $\gamma$ decay energy
yield a weighted average value of 8381.8(24) keV for IAS in $^{29}$P,
which is adopted by NUBASE$^\prime$16~\cite{AME2016}.
There is a discrepancy with respect to the excitation energy for the IAS in $^{29}$Si,
where two values are given at 8290(5) keV \cite{29Si1} and 8310(10) keV \cite{29Si2}.
The excitation energies of $^{29}$Si levels near 8000~keV in the latter reference are about 13~keV
higher than the ones in the former reference.
However, the results on the energy levels in the former reference are more precise and consistent with previous measurements in Refs.~\cite{29Si3,29Si4}.
Here we used the value of 8290(5)~keV for the IAS excitation energy in $^{29}$Si.
The NUBASE$^\prime$16 \cite{AME2016} and NNDC \cite{NNDC} also adopt this value.
According to the quadratic IMME, $ME(^{29}\rm {S})=-3106(17)$~keV is expected,
which is in excellent agreement with our measurement.

By using our new $ME(^{29}\rm {S})$, we have fitted the $ME$ values of the $A=29$, $T=3/2$ isobaric quartet with the quadratic form of IMME.
The results are presented in Table~\ref{IMME table}. 
The obtained $\chi_{n}=0.44$ indicates that the $ME$ values can be well described by the quadratic form of IMME.
We note, that $\chi_{n}= 0.96$ is obtained if the previous $ME(^{29}\rm {S})=-3160(50)$ keV is used.

\begin{table}[t]
	\caption{
	Compilation of $ME$ values for ground states (g.s.),
	IASs and the corresponding excitation energies (\emph{E$_{x}$}) of \emph{A} = 29, $J^\pi=5/2^+$, \emph{T} = 3/2 quartet.
	The data are from AME$^{\prime}$16 and NUBASE$^\prime$16 \cite{AME2016} except for $^{29}$S. 
	Also listed are the parameters of the quadratic and cubic IMME fits (see text).}
	\begin{tabular*}{8.65cm}{lcccc}
		\hline
		Atom & $T_{z}$  & $ME$(g.s.) & $E_{x}$ & $ME$(IAS) \\
		     &          & (keV)      & (keV)& (keV) \\
		\hline
		$^{29}$S   & $-3/2$  & $-3094(13)^{*}$~& 0  & $-3094(13)$ \\
		$^{29}$P  & $-1/2$  & $-16952.8(4) $ & $8381.8(24)$ & $-8571.0(25)$ \\
		$^{29}$Si  & $+1/2$  & $-21895.0784(6)$ & $8290(5)$ & $-13605(5)$ \\
		$^{29}$Al  & $+3/2$  & $-18207.8(3)$    & 0    & $-18207.8(3)$ \\
		\multicolumn{3}{l}{Quadratic fit: $\chi$$_{n} = 0.44$} & & \\
		&  & $a$ (keV)    & $b$ (keV) & $ c $ (keV) \\
		&  & $-11143.2(31)$  & $-5036.5(34)$ & $ 217.8(33) $  \\
		\multicolumn{3}{l}{Cubic fit: $d = -2.0(35)$ keV} & & \\
		&  & $a$ (keV)    & $b$ (keV) & $ c $ (keV) \\
		&  & $-11142.6(32)$  & $-5033.5(63)$ & $ 218.6(35) $  \\
		\hline
		\multicolumn{3}{l}{$^{*}$ this work.}\\
	\end{tabular*}
	\label{IMME table}
\end{table}
\begin{figure}[b]
	\includegraphics[width=0.9\columnwidth]{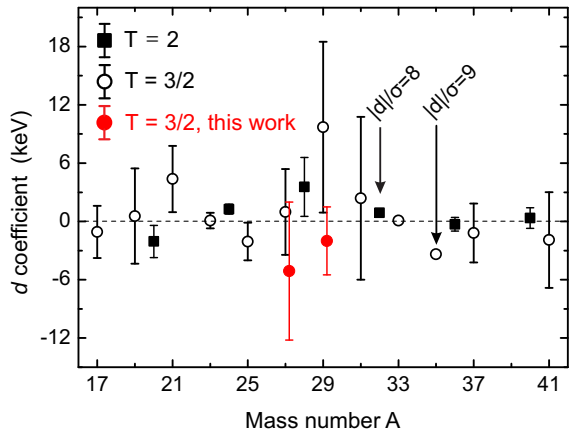}
	\caption{(Color online). $d$ coefficients of the cubic terms in IMME for the isospin quartets and quintets in the $sd$-shell. Data are taken from Ref.~\cite{2014MA56}. Both values, the previously known as well as those obtained in this work, are shown for $A=27$ and 29. }
	\label{d-coef}
\end{figure}
\par

We have applied the cubic form of IMME to describe the $ME$ values of the $A=29$ isobaric quartet.
The obtained parameters are listed in Table~\ref{IMME table}.
The $a$, $b$, $c$ parameters are in good agreement with those obtained from the quadratic fit,
and the $d$-coefficient, $-2.0\pm 3.5$ keV, is consistent with zero at $1\sigma$ confidence level.
We note, that $d=9.0\pm 8.8$ keV is obtained if the former $ME(^{29}\rm {S})=-3160(50)$ keV~\cite{AME2016} is used.

Figure~\ref{d-coef} shows a compilation of the $d$ coefficients~\cite{2014MA56} for the isobaric quartets and quintets in the $sd$-shell.
The absolute $d$ coefficients are typically smaller than about 5 keV.
Even for the considered significant breakdowns of the quadratic IMME, e.g., for the $A=32$ and $A=35$ multiplets,
the absolute $d$ values are still less than 5 keV.
By using our mass excess values of $^{27}$P and $^{29}$S, the extracted $d$ values are well within this limit.


As has been discussed in Ref.~\cite{Bentley07}, the $b$ and $c$ coefficients
are related to the isovector and isotensor components of the isospin non-conserving interactions, respectively.
Their values
can yield the information on the charge symmetry and charge independence of the attractive nucleon-nucleon interaction.
The $b$ and $c$ values have been compiled in Ref.~\cite{2014MA56}.
Removing the known gross $A$ dependence of the $b$ and $c$ coefficients, a more detailed picture is exhibited by plotting
the so-called Coulomb radius parameters defined as~\cite{1979IMME}:
\begin{eqnarray}\label{r-zero}
r_{0b}=\frac{3}{5}\frac{e^2(1-A)}{(b-\Delta_{nH})A^{1/3}}~,~~~ r_{0c}=\frac{3}{5}\frac{e^2}{c~A^{1/3}}~,
\end{eqnarray}
where $\Delta_{nH}=782.3$ keV is the hydrogen-neutron mass difference.
The ratio of the $r_{0b}$ and $r_{0c}$ radii, $R$, was defined in Ref.~\cite{1979IMME} through the $b$ and $c$ coefficients of IMME:
\begin{eqnarray}\label{Ratio}
R=\frac{b-\Delta _{nH}}{c(1-A)}~~.
\end{eqnarray}
The $r_{0b}$ and $r_{0c}$ radii have different nature and
can respectively be associated with averaged properties of the core and valence particles~\cite{1979IMME}.
In the simple assumption of the rigid, homogeneously charged sphere, $R=r_{0c}/r_{0b}=1$.
On the one hand, the quantity $R$ reflects the interplay of isovector and isotensor components of the isospin non-conserving interactions.
On the other hand, it is obtained directly from the experimental masses of the $T=3/2$ isospin multiplets.

\begin{figure}[t]
	\includegraphics[width=0.9\columnwidth]{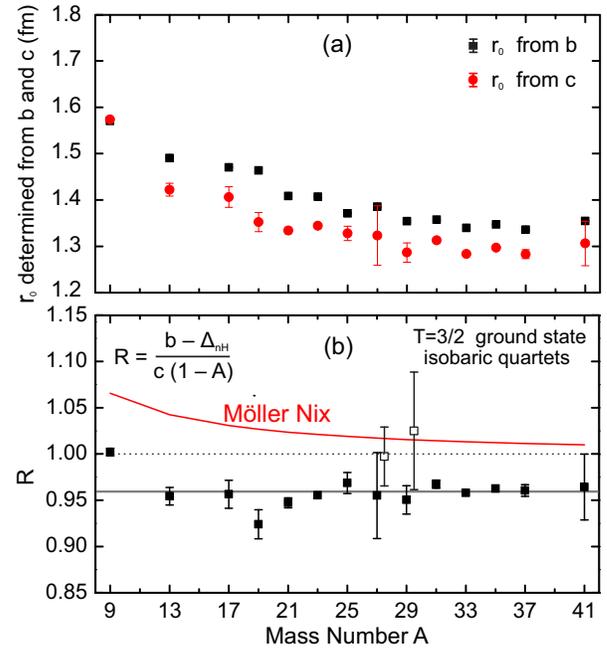}
	\caption{(Color online)
		Top: Coulomb radius parameters determined from the $b$ and $c$ coefficients
		of the cubic IMME
		vs mass number $A$.
		Bottom: The quantity $R$ plotted vs $A$ for ground-state quartets.
		The previously known $R$-values for $A=27$ and 29 are shown with open symbols.
		For a uniform sphere $R=1$.
		The red solid line is derived using the expressions in Ref.~\cite{2013LYH} for the liquid-droplet model~\cite{1988Nix}.
		The black line at $R=0.96$ is to guide the eye.
		}
	\label{Radius}
\end{figure}

\par
The radius parameters $r_{0b}$ and  $r_{0c}$, and the $R$ values are calculated according to Eqs.~(\ref{r-zero}) and (\ref{Ratio})
by using the $b$ and $c$ coefficients from the cubic fits in Ref.~\cite{2014MA56} (see Fig.~\ref{Radius}).
In case the $d$ coefficient equals exactly zero, there is no difference between the results of the quadratic and cubic fits.
For the $A=21,27,29,31$ isobaric multiplets, to determine the corresponding $b$ and $c$ coefficients we have used more recent masses of
$^{21}$Mg~\cite{2014Mg21}, $^{29}$S (this work), $^{31}$Cl~\cite{2016KA15}, and $^{27}$P (this work)
and the energies of the associated  IAS's from Ref.~\cite{AME2016}.

The results for $A=11$, 15 and 39 multiplets are omitted here.
For $A=11$, instabilities with the $b$ and $c$ values between quadratic and cubic fits were observed in Ref.~\cite{2014MA56}.
In the case of the $A=15$ and 39 multiplets there are no cubic fits presently possible.
The $R$-value extracted from the quadratic fit for $A=15$ multiplet is significantly larger than unity
and for $A=39$ it is compatible with unity.

Early investigations have shown that the experimentally deduced $R$ values were fluctuating around unity, and no definite difference between $r_{0b}$ and $r_{0c}$ was observed~\cite{1979IMME}.
This was mainly due to large mass uncertainties for the $T_z=-3/2$ exotic nuclei, especially, for $^{27}$P, $^{29}$S, and $^{31}$Cl.
Most of these uncertainties have been removed through precision mass measurements in the recent years.
It is of interest to conclude from Fig.~\ref{Radius} that $r_{0c}$ are definitely lower than $r_{0b}$, and all $R$ values fall to a constant
value of about $0.96$ with only exception of the $A=9$ multiplet.
We note, that the scatter of $R$-values obtained from quadratic fits is in general larger than of those from cubic fits illustrated in Fig.~\ref{Radius}.

Under the simplified assumptions of an uniformly charged sphere and of Coulomb interaction as the only source of the breaking of the isospin symmetry, $R$ is related to the charge radius of an isobaric multiplet via $R=r_{0c}/r_{0b}$.
The deviation of $R$ from unity reflects the incompleteness of this assumption and suggests the inevitable role of nuclear interaction in breaking the isospin symmetry.
The apparently lower $r_{0c}$, as illustrated in Fig.~\ref{Radius}, shows that the valence particles have greater $rms$ radius than the core~\cite{1979IMME}.
The latter can affect the extraction of nuclear form factors from Coulomb displacement energies (see, e.g., Ref.~\cite{NS}).

The $b$ and $c$ coefficients contain information on isospin-symmetry breaking originating not only from the Coulomb interaction but also from the attractive nucleon-nucleon interaction~\cite{Bentley07}.
It is striking, that the theoretical expressions for $b$ and $c$ coefficients~\cite{2013LYH} derived for the liquid droplet model by M\"{o}ller and Nix~\cite{1988Nix} fail to reproduce the experimental results (see Fig.~\ref{Radius}).
Furthermore, in current shell model calculations, such additional interactions \cite{SY1,SY2,SY3} are added to an isospin-conserving Hamiltonian, with the isovector and isotensor components empirically fitted to data.
Thus, the nearly constant $R\approx0.96$
deduced from the data provides a sensitive experimental constraint for theoretical calculations.

The confirmation of a constant $R$ value can be used for critical checks of the reliability of the measured masses.
Our experimental masses and the analyses have already shown that the 66-keV mass difference for $^{29}$S leads to a drop of $R$ value from 1.02(6) to 0.95(2).
A similar drop of $R$ value from 1.00(3) to 0.96(5) is also found for $A=27$ quartet if our new mass value for $^{27}$P is used.
It is interesting to examine the $fp$-shell nuclei and/or higher isospin multiplets whether or not the ratio $R$ remains constant.

In summary, by using the IMS technique in CSRe, we measured the mass of $^{29}$S.
The new mass excess, $ME$($^{29}$S) $=-3094(13)$, is 66(52)~keV larger than the tabulated value.
The uncertainty is decreased by a factor of 3.8.
The new $ME$ value of $^{29}$S, together with those of the $T=3/2$ isobaric analog states (IAS) in $^{29}$P, $^{29}$Si, and $^{29}$Al,
fit well into the quadratic form of the Isobaric Multiplet Mass Equation, IMME.
Also we remeasured the mass of $^{27}$P, $ME(^{27}$P$)=-685(42)$ keV.

We have analyzed the $b$ and $c$ coefficients of the IMME for the $sd$-shell $T=3/2$ isospin multiplets
and found that the ratio of Coulomb radius parameters introduced by Beneson and Kasy in Ref. \cite{1979IMME}, $R$,
is nearly constant for all considered multiplets with $A>9$, $R\approx0.96$.
Such a constant
parameter may be used as a sensitive test of theoretical calculations.
On the one hand, the occurrence of the near-constant $R$ value for all the nuclei shown in Fig.~\ref{Radius} is striking.
It may imply that the isospin non-conserving forces come from a deeper origin,
same for all nuclei and independent of detailed structure of individual $A$-multiplets.
It is particularly interesting when one notices the fact that some $T_z=-3/2$ nuclei involved in this study
are either unbound or very-loosely-bound (such as $^{19}$Na, $^{23}$Al, $^{31}$Cl, $^{35}$K),
while others are well-bound, between which one may expect large structure differences due to the geometrical effect \cite{YS1}.
Furthermore, the understanding of the near-constant $R$ value may be useful for resolving the long-standing Nolen-Schiffer anomaly
in calculations of Coulomb displacement energies \cite{NS}.
On the other hand, the near-constant $R$ value may be used to predict still unknown masses of neutron-deficient nuclei.

In order to confirm and to further constrain the constancy of the $R$-value, the uncertainties of individual data points need to be reduced, see Fig. \ref{Radius}(b).
In particular we suggest to re-measure the mass of $^{19}$Na nucleus, which is responsible for about 2.3$\sigma$ deviation from $R\approx0.96$.
Evidently, it is interesting to extend such studies to the $fp$-shell nuclei and/or the higher isospin multiplets both in theory and experiment.


\acknowledgments
We thank the staffs of the accelerator division of the IMP for providing stable beam.
Fruitful discussions with Jason Holt, Achim Schwenk and Ragnar Stroberg are greatly acknowledged.
This work is supported in part by the Key Research Program of Frontier Sciences of CAS (Grant No. QYZDJ-SSW-S),
the National Key Program for S\&T Research and Development (Grant No. 2016YFA0400504),
the Helmholtz-CAS Joint Research Group HCJRG-108,
and
the European Research Council (ERC) under the European Union's Horizon 2020 research and innovation programme (grant agreement No 682841 ``ASTRUm'').
Y.A.L. acknowledges Chinese Academy of Sciences for support through the CAS visiting professorship for senior international scientists (Grant No. 2009J2-23) and the CAS External Cooperation Program (Grant No. GJHZ1305).
K.B. acknowledges support by the Nuclear Astrophysics Virtual Institute (NAVI) of the Helmholtz Association.
Y.H.L. thanks the support from the Ministry of Science and Technology of China (Talented Young Scientist Program)
and from the China Postdoctoral Science Foundation (2014M562481).
X.X. thanks the support from CAS "Light of West China" Program.
Y.H.Z. acknowledges support by the ExtreMe Matter Institute EMMI at the GSI Helmholtzzentrum f{\"u}r Schwerionenforschung, Darmstadt, Germany.
X.L.T. thanks the support from the Max Planck Society through the Max-Planck Partner Group.

\end{document}